\documentstyle[aaspp4,11pt]{article} 

\def\gsim{\;\rlap{\lower 2.5pt
\hbox{$\sim$}}\raise 1.5pt\hbox{$>$}\;}
\def\lsim{\;\rlap{\lower 2.5pt
   \hbox{$\sim$}}\raise 1.5pt\hbox{$<$}\;}

\newcommand{\be}	{\begin{equation}}
\newcommand{\ee}	{\end{equation}}
\newcommand{\bd}	{\begin{displaymath}}
\newcommand{\ed}	{\end{displaymath}}

\begin{document}

\title{Low--Mass Proto--Planet Migration in T-Tauri $\alpha$--Disks}

\author{Kristen Menou\altaffilmark{1,2}}
\affil{Virginia Institute of Theoretical Astronomy, Department of Astronomy,}
\affil{University of Virginia, Charlottesville, VA 22903, USA}

\and

\author{Jeremy Goodman}
\affil{Department of Astrophysical Sciences, Princeton University
Princeton, NJ 08544, USA}

\altaffiltext{1}{Celerity Foundation Fellow} 
\altaffiltext{2}{Current Address: Institut d'Astrophysique de Paris,
98bis Boulevard Arago, 75014 Paris, France}

\begin{abstract}
We present detailed estimates of ``type-I'' migration rates for
low-mass proto-planets embedded in steady-state T-Tauri
$\alpha$--disks, based on Lindblad torque calculations ignoring
feedback on the disk.  Differences in migration rates for several plausible
background disk models are explored and we contrast results obtained
using the standard two dimensional formalism of spiral density wave
theory with those obtained from a simple treatment of
three-dimensional effects.  Opacity transitions in the disk result in
sudden radial variations of the migration rates. Regions with minimal
migration rates may be preferred sites of gravitational interactions
between proto--planets. Three-dimensional torques are significantly
weaker than two-dimensional ones and they are sensitive to the surface
density profile of the background disk. We find that migration times
in excess of runaway envelope accretion times or T-Tauri disk
lifetimes are possible for Earth-mass proto-planets in some background
disk models, even at sub--AU distances. We conclude that an
understanding of the background disk structure and ``viscosity'', as
well as a proper treatment of three-dimensional effects in torque
calculations, are necessary to obtain reliable estimates of ``type-I''
migration rates.


\end{abstract}

\section{Introduction}

Since the first discovery of a planet orbiting a nearby Sun-like star
about a decade ago (Mayor \& Queloz 1995), radial velocity surveys of
a few thousands of our closest stellar neighbors have uncovered more
than a hundred such planets (Marcy \& Butler 1998; Cumming, Marcy \&
Butler 1999; Marcy, Cochran \& Mayor 2000). Transit searches, which
have confirmed the gaseous giant nature of these planetary companions
(Charbonneau et al. 2000), promise to uncover a large number of
additional extrasolar planets in the future (Udalski et al. 2002;
Konacki et al. 2003).

In addition to the large eccentricities of many of the planets
discovered to date, one of the most striking features of this new
population is the existence of a subset of extrasolar planets orbiting
very close to their parent star (see, e.g., the extrasolar planet
almanac\footnote{{\tt http://exoplanets.org/almanacframe.html}} and
encyclopedia\footnote{{\tt http://www.obspm.fr/encycl/encycl.html}}
for a census). It is generally accepted that these close-in extrasolar
giant planets could not have formed {\it in situ}, but instead must have
migrated from large distances, where conditions for planet formation
are more favorable (see, e.g., Lin, Bodenheimer \& Richardson 1996).

There are two physically distinct scenarios considered for the
formation of gaseous giant planets (see, e.g., Ruden 1999; Wuchterl,
Guillot \& Lissauer 2000 for reviews). In the standard core accretion
scenario, dust grain accumulation in a proto-planetary disk proceeds
to the formation of planetesimals, which later assemble to form
proto-planetary cores. Above a mass threshold of a few tens $M_\oplus
$, runaway accretion of the surrounding gas onto the proto-planetary
cores occurs, leading ultimately to the formation of giant planets
made of gas for the most part (Safronov 1969; Mizuno 1980; Pollack et
al. 1996; see also Rafikov 2003). Cores of giant planets form more easily
beyond the "snow line," whose location is typically at AU distances
from the parent star (Hayashi 1991; Boss 1996; Sasselov \& Lecar
2000). In the alternative disk instability scenario, a rather massive
proto-planetary disk is subject to gravitational instabilities leading
to the direct formation of (possibly multiple) Jupiter-sized objects
(Cameron 1978; Boss 1997; 2000, Armitage \& Hansen 1999; Mayer et
al. 2002; Lufkin et al. 2003). In the present study, our interests are
more focused on the core accretion scenario because the evolution of
embedded proto-planets is of general interest, whether or not
gravitational instabilities are important for giant planet formation.

In both planet formation scenarios, inward migration is required to
explain the existence of the population of close-in extrasolar giant
planets. Here again, two physically distinct scenarios for migration
have been put forward. Spiral density waves launched by a proto-planet
embedded in a gaseous disk apply a negative torque on the
proto-planet, which migrates inward as a result of orbital angular
momentum losses (Goldreich \& Tremaine 1980, henceforth GT80; Lin \&
Papaloizou 1986; Korycansky \& Pollack 1993, henceforth KP93;
Artymowicz 1993a,b; Ward 1997a). On the other hand, later in the
system's evolution, it is also possible for a planet embedded in a
sufficiently massive planetesimal disk to experience inward migration
via repeated planetesimal scattering events (Murray et al. 1998). Our
interest in the present study is in the gaseous migration mechanism,
partly because it surely affects the precursors of gas giants, at
least within the core-instability scenario, whether subsequent
planetesimal scattering migration occurs or not.

It has long been recognized that the standard core accretion and
gaseous migration scenarios face a serious difficulty when
combined. While estimated migration times for massive proto-planetary
cores located at $\sim$~AU distances from their parent star are $\sim
10^5$~years (see, e.g., Ward 1997a; KP93), estimates for the time
required to accrete a gaseous envelope are typically in the range
$\sim 10^6$--$10^7$~years (see, e.g., Pollack et al. 1996). Short
inward migration times imply that cores would be accreted by the
central star before they could build up any substantial gaseous
envelope, and more generally, they pose a serious threat to the
survival of any planetary system in formation (see, e.g., Ward 1997b).

Robust calculations for proto-planetary core migration times are
notoriously difficult to achieve, however. While a forming planet with
a mass $\gsim 10$--$100 M_\oplus$ may open a gap and see its
subsequent migration tied to the slow viscous evolution of the gaseous
disk (``type--II'' migration), a lower-mass proto-planetary core
migrates in, relative to the gaseous component (``type--I''
migration), at a rate which strongly depends on the exact disk
conditions in the vicinity of the proto-planet. In that respect, most
calculations of ``type--I'' migration rates to date were made with
rather idealized models for the background disk.  Both linear torque
calculations (Ward 1997a; KP93) and fully non--linear hydrodynamical
simulations (Kley, D'Angelo \& Henning 2001; Nelson \& Benz 2003a,b;
D'Angelo, Kley \& Henning 2003; Lufkin et al. 2003) usually assume
power law profiles for the disk properties, in models often
reproducing the minimum mass solar nebula (MMSN) characteristics. This
is one of our main motivations to study ``type--I'' gaseous migration
with more realistic proto-planetary disk models. While there are
significant uncertainties regarding the nature of angular momentum
transport in these disks (Gammie 1996; Glassgold, Najita \& Igea 1997;
Fromang, Terquem \& Balbus 2002; Fleming \& Stone 2003; Matsumura \&
Pudritz 2003) or their overall structure (e.g., Chiang \& Goldreich
1997), we believe that a study of how these uncertainties may affect
migration time estimates is useful at this time.

Another source of uncertainty in calculations of ``type--I'' migration
rates comes from limitations of spiral density wave theory itself. The
theory has been initially developed for infinitely-thin,
two-dimensional disks (Goldreich \& Tremaine 1980), but
three-dimensional effects are important (Tanaka, Takeuchi \& Ward
2002). Finite eccentricities and inclinations of embedded
proto-planets (Papaloizou \& Larwood 2000; Goldreich \& Sari 2003;
Ogilvie \& Lubow 2003) and the potentially magnetized (Terquem 2003)
and turbulent nature (Winters, Hawley \& Balbus 2003; Papaloizou \&
Nelson 2003; Nelson \& Papaloizou 2003a,b; Papaloizou, Nelson \&
Snellgrove 2003) of the proto-planetary disk are also ignored in the
original theory, while they could all have important consequences. It
is probably fair to say that a consensus has yet to be achieved on
these various aspects of spiral density wave theory.

Nevertheless, an important motivation for developing a detailed
numerical tool to study gaseous migration in proto-planetary disks is
the intimate relation, which has become increasingly obvious, between
planetary formation and migration on the one hand, and disk evolution
on the other hand. Planets form from disk material, migrate by
interacting with the disk and at the same time influence strongly the
disk long-term evolution by raising torques which affect the disk
global angular momentum budget and eventually open gaps in an
otherwise smooth gaseous distribution. Therefore, it is likely that a
general planet formation theory capable of explaining the diversity of
known systems will require following the strongly-coupled evolution of
a potentially large ensemble of proto-planets together with the disk
in which they formed {(see, e.g., Lin \& Papaloizou 1986 for early
calculations of this type)}.

In this first investigation of gaseous migration in proto-planetary
disks, we focus on ``type--I'' migration of low-mass proto-planets,
ignoring the feedback from Lindblad torques on the disk structure.
The radial structure of our disks is not a simple power law but
depends upon opacities and other details of accretion. We explore how
uncertainties in the structure of proto-planetary disks may affect
``type--I'' migration rates and how sensitive these rates are to
three-dimensional effects. In \S2, we describe the characteristics of
our T-Tauri disk model, while in \S3 the two- and three-dimensional
formulations we have adopted for Lindblad torques are summarized. {
Corotation torques, and our reasons for neglecting them, are discussed
briefly in \S3.3.}  We present our results in \S4, we discuss possible
limitations and extensions of this work in \S5 before concluding in
\S6.

\section{Disk Models}

The present study is exclusively focused on steady-state
$\alpha$--disk models (see \S4.1 for some justification). In the
future, however, we intend to study the coupled evolution of
planet-disk systems with explicit planetary migration, and this
obviously requires being able to follow the evolution of the disk's
properties with time. With this in mind, we have developed a fully
time-dependent model for T-Tauri $\alpha$--disks. It is largely
inspired by the model developed by Hameury et al. (1998) to study
the time-dependent evolution of disks around compact objects. The main
assumption in our models is that disks are geometrically thin, so that
their radial and vertical structures can effectively be treated
separately.

\subsection{Radial Structure and Evolution}

Following Hameury et al. (1998), we solve the following radial
equations for the conservation of mass, angular momentum, and energy:
\begin{eqnarray}
{\partial \Sigma \over \partial t} & = & - {1 \over r} {\partial \over
\partial r} (r \Sigma v_{\rm r}) + \dot \Sigma,\\
j{\partial \Sigma \over \partial t} & = & - {1 \over r} {\partial \over \partial
r} (r \Sigma j v_{\rm r}) + {1 \over r} {\partial \over \partial r}
\left(- {3 \over 2} r^2 \Sigma \nu \Omega_{\rm K} \right) + \dot J, \\
{\partial T_{\rm c} \over \partial t} &=& { 2 (Q^ + -Q^- + H) \over C_P
\Sigma} - {\Re T_{\rm c}\over \mu C_P} {1 \over r} {\partial (r v_{\rm
r}) \over \partial r} - v_{\rm r} {\partial T_{\rm c} \over \partial
r},
\end{eqnarray}
where $\Sigma$ is the surface column density, $v_{\rm r}$ is the gas
radial velocity in the disk, $j = (GM_* r)^{1/2}$ is the specific
angular momentum of material at radius $r$ in the disk,
$\Omega_K=(GM_* /r^3)^{1/2}$ is the Keplerian angular velocity, $M_*$
is the mass of the central star (taken to be $0.5 M_\odot$ in the
present work), $T_c$ is the mid-plane temperature, $C_P$ is the heat
capacity at constant pressure, $\Re$ is the perfect gas constant,
$\mu$ is the mean molecular weight and $\nu$ is the kinematic
viscosity coefficient, for which we adopt a standard
$\alpha$-prescription (Shakura \& Sunyaev 1973). Justifications for
the use of an $\alpha$--viscosity to describe angular momentum
transport in thin accretion disks have been given by Balbus \&
Papaloizou (1999) in the case of magnetized turbulent transport and by
Goodman \& Rafikov (2001) in the case of transport by
locally-dissipated planetary torques.  The gas motion is implicitly
described as Keplerian in our formulation of the disk
equations. Deviations from strict Keplerianity due to finite radial
pressure gradients are calculated separately, when required for the
planetary torque calculations, and deviations from Keplerianity due to
disk self-gravity are ignored.

The terms $\dot \Sigma$ and $\dot J$ in the first two equations
represent additional sources/sinks of mass and angular momentum,
respectively, for the disk (both are put to zero in the present
study). $Q^+$ and $Q^-$ are the local heating and cooling rates per
unit surface, respectively.  The term $H$ accounts for the radial
energy flux caused by steep radial gradients, and it is estimated in
the framework of the $\alpha$ parametrization (see Hameury et al. 1998
for details). This general form of the energy equation accounts for
the possibility that the disk is out of thermal equilibrium ($Q^+ \neq
Q^-$) as can occasionally happen in time-dependent studies (see, e.g.,
Menou, Hameury \& Stehle 1999 for specific examples in dwarf nova
disks). For all the steady-state models presented here, however, the
radial energy equation reduces to a very good approximation to $Q^+ =
Q^-$.

Equations~(1) and~(2) can be combined to obtain the standard viscous
diffusion equation for a geometrically thin disk (when $\dot \Sigma
=\dot J=0$; see, e.g., Pringle 1981). The above set of equations is
solved on an adaptive numerical grid, to guarantee that regions of the
disk characterized by sharp gradients receive a large number of grid
points and thus adequate numerical resolution. A number $N=400$ grid
points has been adopted for all the calculations presented here. The
equations are evolved with a fully implicit numerical scheme and a
variable time-step. Further details can be found in Hameury et
al. (1998). For the present study, steady-state disk models were
obtained in the limit $\partial /\partial t \to 0$ (in practice, a
large enough time-step).

This set of partial differential equations requires a total of six
boundary conditions. For a disk with fixed inner and outer radii which
is fed, at its outer edge, at a rate $\dot M = \dot M_{\rm out}$, the
boundary conditions at the inner edge ($n=1$) and the outer edge
($n=N$) of the numerical domain are:
\begin{eqnarray}
\Sigma | _{n=1} &=& \epsilon \ll 1~{\rm g~cm}^{-2} , \nonumber \\
r | _{n=1} &=& r_{\rm in} ~(0.01~AU), \nonumber \\
\frac{\partial T_c}{\partial r} | _{n=1} & =& 0, \nonumber\\
\dot M | _{n=N} &=& \dot M_{\rm out}, \nonumber \\
r | _{n=N} &=& r_{\rm out} ~(200~AU), \nonumber \\
\frac{\partial T_c}{\partial r} | _{n=N} & =& 0. 
\end{eqnarray}
The precise value of $\Sigma | _{n=1}$ is unimportant as long as it is
small compared to values in the bulk of the disk.

\subsection{Vertical Energy Transport}

Contrary to Hameury et al. (1998), we do not explicitly solve for
radiative-convective energy transport in the vertical
direction. Instead, we calculate the local cooling rate $Q^- \equiv 2
\sigma T_{\rm eff}^4$ (for 2 disk faces) with a one-layer version of
the radiative transfer model of Hubeny (1990). According to this
model, the disk intrinsic effective temperature, $T_{\rm eff}$, is
related to the disk mid-plane temperature, $T_c$, by

\begin{equation}
T_{\rm eff}^4 = \frac{4}{3} \frac{\left( T_c^4 - T_{\rm irr}^4
\right)}{\left( \tau_{\rm ext}/2 + 1/\sqrt{3}+1/(3 \tau_{\rm abs})
\right)},
\label{eq:hubeny}
\end{equation}
where the mid-plane optical depth $\tau=\kappa \Sigma /2$,
$\kappa_{\rm ext}$ is the Rosseland-mean extinction opacity
(absorption+scattering), $\kappa_{\rm abs}$ is the Rosseland-mean
absorption opacity and $T_{\rm irr} > 0$ in the presence of external
irradiation.  Note that in this formulation, $T_{\rm eff}$ accounts
only for the disk intrinsic emission (i.e. from viscous dissipation
only), while the disk also re-emits all the irradiation flux absorbed
locally (as measured by $T_{\rm irr}$). A detailed discussion of this
radiative transfer model and of how well it does in the two important
optically-thick and optically-thin limits can be found in Hubeny
(1990).  While Hubeny's model requires in principle a Planck-mean
opacity to be used for $\tau_{\rm abs}$ (in the optically-thin
regime), we only use Rosseland-mean opacities here (see \S2.3 for a
justification). We do not allow for different opacities when
considering the disk's own radiation and the hotter radiation from the
central star. Adopting this simple prescription for vertical radiative
transfer is an oversimplification of the problem, which is justified
{\it a posteriori} by satisfactory comparisons to previous studies
treating the vertical energy transport problem in more details.  Our
one-layer approximation neglects vertical transport by convection,
whose role is unclear, especially in the context of magnetized
accretion disks, which are already subject to a dynamical instability
leading to turbulent transport, by construction.  We also note that
D'Alessio et al. (1998) have estimated that convective transport
should be unimportant in all but the very innermost regions of a
typical T-Tauri disk.

Viscous dissipation is prescribed in the usual form, with a heating
rate per unit area
\begin{equation}
Q^+ = \frac{9}{4} \nu \Sigma \Omega_K^2,
\end{equation}
where the kinematic viscosity, $\nu = \alpha c_s h$, is expressed in
terms of a dimensionless viscosity parameter, $\alpha$ ($c_s$ is the
local adiabatic sound speed and $h$ is the local scale height).

Irradiation by the central star is an important source of heating for
the disk at large distances. Rather than solving self-consistently for
the magnitude of irradiation based on a detailed model of the disk
shape and structure, as in Dubus et al. (1999), we parametrize it with
a simple geometrical model. Assuming that the irradiation source is
point-like (a reasonable assumption far away from the stellar source),
the irradiation flux is given by
\begin{equation} \label{eq:irrflux}
\sigma T_{\rm irr}^4=\frac{L_* (1-\epsilon)}{4 \pi  r^2} \frac{h}{r}
\left( \frac{d \log h}{d \log r} - 1 \right),
\end{equation}
where $\sigma$ is the Stefan-Boltzmann constant, $L_*$ is the stellar
luminosity, $\epsilon$ is the disk albedo and $T_{\rm irr}$ should go
to zero if the last term in parenthesis becomes negative (disk
self-shadowing; see, e.g., Dubus et al. 1999). For $\epsilon =0.5$, an
aspect ratio $h/r \sim 0.1$, a geometric factor $\sim 0.1$ (last term
in parenthesis in Eq.~[\ref{eq:irrflux}]) and $L_* \sim L_\odot$, this
yields an irradiation temperature
\begin{equation} \label{eq:tirr}
T_{\rm irr} \simeq 90~K~\left( \frac{r}{1~{\rm AU}} \right)^{-1/2},
\end{equation}

{which is directly used in equation~(\ref{eq:hubeny}).  This
prescription artificially exposes disk regions with $d \log h / d \log
r < 1$, that would normally be self-shadowed, to irradiation from the
central star, but without serious consequences as we shall see below.
Although an improved treatment of disk irradiation would lead to
better accuracy, this simple prescription should be sufficient for our
purposes in view of other large uncertainties inherent to structural
models of proto-planetary disks.}

\subsection{Opacities and Equation of State}

We assume that dust (when present) is fully mixed with the
gas. Analytic expressions for the Rosseland-mean extinction opacities
are taken from Bell et al. (1997; see also Henning \& Stognienko
1996). Pollack et al. (1994) have found that Planck-mean and
Rosseland-mean extinction opacities are comparable for mixed
dust. This justifies our use of Rosseland-mean opacities in
Eq.~(\ref{eq:hubeny}) even in the optically-thin regime. Following
Pollack et al. (1994; see their Fig.~4b), we also adopt $\kappa_{\rm
abs} = \kappa_{\rm ext}$, since we have found that this is adequate
for the range of temperatures at which typical T-Tauri disks are
optically-thin.

The equation of state adopted is that of a perfect gas. The mean
molecular weight and adiabatic index are typically $\mu_{H_2} \simeq
2.4$ and $\Gamma \simeq 7/5$ throughout the disk, since hydrogen is
everywhere predominantly in molecular form. Other thermodynamical
quantities are self-consistently calculated as a function of
temperature and density at the disk mid-plane when needed (see Hameury
et al. 1998 for details).

\section{Planetary Torques}

In this study, we focus on the low-mass perturber limit and we assume
that the feedback on the disk due to the presence of embedded
proto-planets can be neglected. We neglect corotation torques in our
calculations, and assume for simplicity that all the planets are on
circular orbits, without inclination relative to the disk
mid-plane. Linear spiral density wave theory has been elaborated by a
number of authors (GT80; KP93; Artymowicz 1993a,b; Ward 1997a) and we
rely heavily on these works in our formulation of the problem. Thus,
we calculate the ``torque density'' representing waves excited at
Lindblad resonances, which are located at radii $r_m$ where the disk's
angular velocity $\Omega(r)$ and epicyclic frequency $\kappa(r)$ are
related to the angular velocity $\Omega_p$ of the planet by
$m\left[\Omega(r_m)-\Omega_p\right] =\kappa(r_m)$ for azimuthal
harmonics $m$.

\subsection{Torque Density}

Following Ward (1997a), the torque density (torque per unit radius) of
waves excited in the disk by a planet of mass $M_p=\mu M_*$ orbiting a
star of mass $M_*$ at radius $r_p$ is

\be\label{DTLR}
\left[ \frac{dT}{dr}\left( r \right) \right]_{LR}=\mbox{sign}(r-r_p) 
\frac{2 \mu^2\Sigma
r_p^4\Omega_p^4} {r (1+4\xi^2)\kappa^2}\,m^4\,\psi^2.  
\ee 

The surface density, $\Sigma$, and epicyclic frequency, $\kappa$ (not to be confused with opacity), are
evaluated at the radius of interest, $r$, and $\xi\equiv
mc_s/r\kappa$, where $c_s$ is the adiabatic sound speed.  This
expression incorporates the torque cutoff \emph{via} the functional
dependence of $\xi(r)$, $\psi(r)$ and $m(r)$ (see below).

Note that
\be
1+\xi^2 = \frac{(\Omega-\Omega_p)^2}{(\Omega-\Omega_p)^2~-c_s^2/r^2}.
\ee

The angular velocity and epicyclic frequency in the gaseous disk are
slightly different from their Keplerian values due to finite radial
pressure gradients:

\begin{eqnarray}\label{OMEGA}
\Omega^2(r)&=& \frac{GM_*}{r^3}
+ \frac{1}{r \rho}\frac{\partial}{\partial r}(\rho c_{\rm iso}^2),\\
\kappa^2&\equiv& \frac{1}{r^3}\frac{\partial}{\partial r}(r^4\Omega^2),
\nonumber
\end{eqnarray}

where $c_{\rm iso}$ is the isothermal sound speed.  We must include
the pressure terms in the disk rotation curve when we compute the
torque density, even if we neglect them in the equations for the disk
evolution (\S2.1).  The angular velocity of the satellite, $\Omega_p$,
has the Keplerian value $\sqrt{GM_*/r_p^3}$, and the small difference
$\Omega_p-\Omega(r_p)$ is the most important cause of the imbalance
between the torques exerted on the interior and exterior parts of the
disk, and hence of the planet's migration.  Because of the extra
radial derivative, $\kappa$ is even more sensitive to pressure
effects.\footnote{There is some danger that $\kappa^2$ may be formally
negative near the edges of gaps, which are not considered in the
present study.}

The dimensionless azimuthal wavenumber $m$ is treated as a continuous
function of radius.  It is computed from the WKBJ dispersion relation
by setting $k\to m/r$ (since the radial wavenumber vanishes at the
Lindblad resonances where the waves are excited):

\be\label{MOFR}
m(r) = \left[\frac{\kappa^2}{(\Omega-\Omega_p)^2 - c_s^2/r^2}\right]^{1/2}.
\ee

Where equation~(\ref{MOFR}) predicts imaginary $m$, $dT/dr\to 0$.
Disk self-gravity is neglected in equation~(\ref{MOFR}).

The dimensionless satellite forcing function is
\be\label{PSI}
\psi \equiv \frac{\pi}{2}\left[~\frac{1}{m}\left|\frac{db_{1/2}^m}{d\beta}
\right|~+~2\frac{\Omega}{\kappa}\sqrt{1+\xi^2}~b_{1/2}^m(\beta)\right],
\ee

where $\beta\equiv r/r_p$.  The factor $\Omega/\kappa$ does not appear
in the original formulation of Ward (1997a) but it is needed so that
equation~(\ref{PSI}) reduces to equation~(18) of Goldreich \& Tremaine
(1980), in the limit $\xi\ll 1$.  The standard definition of the
Laplace coefficient is

\be 
b_{1/2}^m(\beta) = \frac{2}{\pi}\int\limits_0^\pi\frac{\cos
m\theta d \theta} {\sqrt{1-2\beta\cos\theta+\beta^2}}.  
\ee

This is exact for an infinitely-thin two-dimensional disk, as studied,
for instance, by Ward (1997a), and this is the formulation we adopt
for 2D Lindblad torques.

To account for three-dimensional effects, we replace the above
expression with the following approximation, which includes a
correction for the thickness, $h$, of the disk

\be\label{LAPM}
b_{1/2}^m(\beta) \approx \frac{2}{\pi\sqrt{\beta}}\,K_0\left(
m\sqrt{\beta-2+\beta^{-1}~+h^2/rr_p}\right),\qquad
h\equiv\frac{c_{\rm iso}}{\Omega}\,,
\ee

where $K_0$ is the modified Bessel function of the second kind of
order $0$: $K_0(z)\approx -\ln z$ for $0<z\ll 1$ and $K_0(z)\approx
e^{-z}\sqrt{\pi/2z}$ for $z\gg 1$. Following Goldreich \& Tremaine
(1980), we have made use of the integral representation \be K_0(z) =
\int\limits_0^\infty\frac{\cos(zt)}{\sqrt{t^2+1}}\,dt \ee (Abramowitz
\& Stegun \S 9.6.21).  The appearance of a term proportional to $h^2$
in equation~(\ref{LAPM}) comes from expressing the gravitational
potential of the perturbing planet as

\begin{equation} \label{eq:pot}
\Phi_p=-\frac{G M_p}{\sqrt{ |{\bf r-r_p}|^2 + h^2}},  
\end{equation}
 while $h$ is set exactly to zero in the two-dimensional theory.

With this expression, we account directly in the perturbing potential
for the fact that the disk material subject to the gravitational
influence of the proto-planet is actually distributed vertically over
a length scale typically $\sim h$, the disk scale height.  It is clear
that this formulation accounts only approximately for
three-dimensional effects. In a detailed study, Tanaka et al. (2002)
find that the exact vertical distribution of mass in the disk enters
explicitly a three-dimensional formulation of spiral density wave
theory (see their Eq.~[28], which is derived under a specific
vertically isothermal assumption). As we emphasized earlier, the
structure of proto-planetary disks, including their vertical
structure, is not well known (e.g., Fromang et al. 2002; Fleming \&
Stone 2003; see also Chiang \& Goldreich 1997). In this context, it is
unlikely that three-dimensional torques can be calculated very
accurately. This is our main motivation for adopting a simple
formulation of three-dimensional effects, that should nonetheless
capture their most salient features. We note that Papaloizou \&
Larwood (2000) have adopted a similar formulation, with a vertical
``softening length'' varying from $40 \%$ to $100 \% $ of a disk
scale-height, $h$ { (see also Ward 1989 for a different treatment
of vertical thickness)}. {It is unclear what value of the softening
length is most appropriate or best reproduces the results of Tanaka et
al. (2002) in general, but most cases of interest should be covered by
a treatment intermediate between our two- and three-dimensional
prescriptions. }

The expression for $\psi$ includes a derivative with respect to
$\beta=r/r_p$, and we include the implicit dependence on $r$
via $h^2(r)/r$ when evaluating this derivative.  However, $m$
is treated as a constant during the differentiation.  To evaluate the
term $db_{1/2}^m(\beta)/d\beta$ in equation~(\ref{PSI}), note that
$dK_0(z)/dz =-K_1(z)$ (Abramowitz \& Stegun \S 9.6.27), where $K_1$ is
the modified Bessel function of the second kind of order $1$.
Therefore 
\be \frac{db_{1/2}^m}{d\beta}(\beta)=
-(2\beta)^{-1}b_{1/2}^m(\beta) ~-\frac{m\,K_1\left(
m\sqrt{\beta-2+\beta^{-1}~+h^2/rr_p}\right)}
{\pi\sqrt{(\beta-1)^2~+h^2/r_p^2}}
\,\left[1-\beta^{-2}+\frac{d}{dr}\left(\frac{h^2}{r}\right)\right]\,.
\ee Since $h=c_{\rm iso}/\Omega$,
\[
\frac{d}{dr}\frac{h^2}{r}= \left(\frac{d}{dr}\ln T_c~+~\frac{2}{r}\right)
\frac{h^2}{r}\,.
\]

When $h\ll |r-r_p|\ll r_p$, then $\xi\ll 1$ and $\psi$ reduces
approximately to a constant:
\bd
\psi\approx K_1(\kappa/|2A|) +\frac{2\Omega}{\kappa}K_0(\kappa/|2A|),
\ed
where $2A\equiv d\Omega/d\ln r=(\kappa^2-4\Omega^2)/2\Omega$ is the
shear rate.  Unfortunately, the torque is dominated by the region
$|r-r_p|\sim h$, where this approximation cannot be used.  In
fact, as $|r-r_p|\to 2h/3$, $m$ diverges and $\psi\to 0$ exponentially.
The torque density vanishes at $|r-r_p|\le 2h/3$: in this region,
the flow past the planet is subsonic so there is no wave drag.

\subsection{Implementation of Planetary Torques in Disk Models}

The torque density has a sharply peaked profile, cutting off at a
distance $\sim h$ from the proto-planet's location (see, e.g., Ward
1997a). An accurate calculation of the torque density therefore
requires that radial profiles in the background disk model be
adequately resolved on length-scales smaller than a disk scale
height. This is not trivially achieved because the typical grid
spacing in our disk models is usually not very much smaller than a
disk scale height.

A first method we explored to calculate the torque density was to make
use of the adaptive grid. In principle, by requiring that the grid
spacing be inversely proportional to the local torque density, one
should reach adequate spatial resolution in the disk model (in the
vicinity of a proto-planet) for the torque density to be calculated
accurately. In practice, we have found that this solution is difficult
to implement and often results in numerical instability.

Instead, we have opted for a sub-grid method. Whenever required for a
torque density calculation, the radial profiles of relevant quantities
in the disk are fitted to cubic splines in the vicinity of the
proto-planet's location. For the torque density calculation itself,
cubic spline interpolations of the disk profiles provide estimates of
disk properties at distances as close to the proto-planet as required
(whose location does not in general coincide with that of a grid point
in the disk model). This implementation effectively decouples the
accuracy of the torque density calculation from the effective
numerical resolution of the background disk model. Generalization to
multiple proto-planets embedded in the disk is straightforward.

\subsection{Corotation Torques}
{ 

In addition to the Lindblad torques, there are torques at co-orbital
corotation resonances.  The latter occur very near the semimajor axis
of the planet, $r_C-r_p\sim O(h^2/r^2)$.  Eccentric planets have
non-co-orbital corotation resonances as well (\emph{e.g.,} GT80).

The theory of corotation resonances is in a less satisfactory state
than that of Lindblad resonances.  GT80 derived an expression for the
corotation torque due to an individual azimuthal harmonic.  In the
co-orbital case and for $m\gg1$, their expression reduces to
\begin{equation}
\Gamma_m^{C,\,\rm GT}\approx -8m K_0(m|1-\beta|)^2
\left(\frac{GM_p}{r_p}\right)^2
\left[\frac{1}{d\Omega/dr}\frac{d}{dr}\left(\frac{\Omega\Sigma}{\kappa^2}
\right)\right]_{r_C}\,,
\label{eq:GTCR}
\end{equation}
where $\beta\equiv r/r_p$ and the sign pertains to the torque on the
planet.  Pressure forces and the disk thickness were neglected;
normally these effects should be unimportant at $m\ll r/h$.  The
dominant contribution to the total torque on an embedded planet,
however, comes from large $m$.  For the Lindblad resonances, GT80
demonstrated that harmonics $m>r/h$ are suppressed by pressure even in
a two-dimensional disk.  This result has been elaborated by later
authors as summarized above.  Analogous ``torque cutoffs'' were
derived for corotation resonances by Ward (1989) using asymptotic
methods, and by KP93 using numerical integration of the linearized
equations.  The latter concluded that the net torque (summed over $m$)
from co-orbital corotation resonances is comparable to the residual of
Lindblad torques---a result that is suggested by simply truncating the
sum over eq.~(\ref{eq:GTCR}) at $m=r/h$.  Since all harmonics are
proportional to the term in square brackets above, however, the
corotation torque is very sensitive to the local gradient in surface
density---even in sign---and to the \emph{third} derivative of the
rotation curve.  Coincidentally, the torque vanishes in a keplerian
disk with the canonical profile of the minimum-mass solar nebula
($\kappa=\Omega\propto\Sigma\propto r^{-3/2}$).  Unfortunately, the
torque cutoffs derived by Ward (1989) and KP93 do not agree
quantitatively, even though both groups used linear theory.  A subset
of KP93's results were confirmed in the 2D limit of a
three-dimensional linear analysis by Tanaka \emph{et al.} (2002), who
treated the special (and relatively tractable) case of disks that are
both radially and vertically isothermal.

Linear theory itself is problematic.  The angular momentum deposited
at Lindblad resonances is carried away by a wave flux.  But the torque
at corotation accumulates in a small amount of disk material unless
spread by viscous diffusion or by differences in the radial drift
rates of the planet and the disk (Ward 1989; Balmforth \& Korycansky
2001; Masset 2001, 2002).  Hence the corotation resonances are prone
to nonlinear saturation.  The larger the gradient of
$\kappa^2/2\Omega\Sigma$ (potential vorticity) at the resonance, the
stronger is the torque in linear theory, but the more sensitive to
saturation.  In a disk with a smooth profile, saturation is not likely
if the planet's mass is much less than both of the gap-opening
criteria cited in \S4.4.

In summary, we have neglected corotation torques for lack of a simple
and reliable parametrization of their effects.
}

\section{Results}

\subsection{T-Tauri $\alpha$-Disk Models}

Although accretion disks in T-Tauri systems are not expected to be in
steady state, a quasi-steady assumption may be a reasonable first
approximation for all but their outermost regions. Indeed, assuming
the disk is freely spreading, the global disk evolution should be
determined by comparatively slow viscous processes in the outer
regions, while the inner regions are able to follow this evolution via
a succession of quasi-steady states (see, e.g., Cannizzo, Lee \&
Goodman 1990; { Ruden \& Pollack 1991}). Using this general
property, Hartmann et al. (1998) have shown that the observational
characteristics of pre-main sequence stars can be successfully
interpreted with freely-spreading accretion disk models in which the
inner regions go through a succession of quasi-steady states with an
accretion rate that is slowly decreasing with time. They infer mass
accretion rates in the range $10^{-7}$--$10^{-9} M_\odot$~yr$^{-1}$,
with typical values $\sim 10^{-8} M_\odot$~yr$^{-1}$ for $10^6$ year
old T-Tauri star systems.

Based on this, our reference T-Tauri disk model is a steady-state
model with a constant accretion rate $\dot M =10^{-8}
M_\odot$~yr$^{-1}$ throughout. We adopt $M_*=0.5 M_\odot$ for the mass
of the central star and a default value of the viscosity parameter
$\alpha=0.02$. Disk irradiation by the central star, when included,
follows the prescription given in equation~(\ref{eq:tirr}). Later on,
we will vary various disk parameters to study how changes in the
background disk model influence the planetary migration process. We
detail here the properties of our reference disk model because it
illustrates rather well the basic characteristics of all the disk
models we have considered.

\begin{figure}
\plottwo{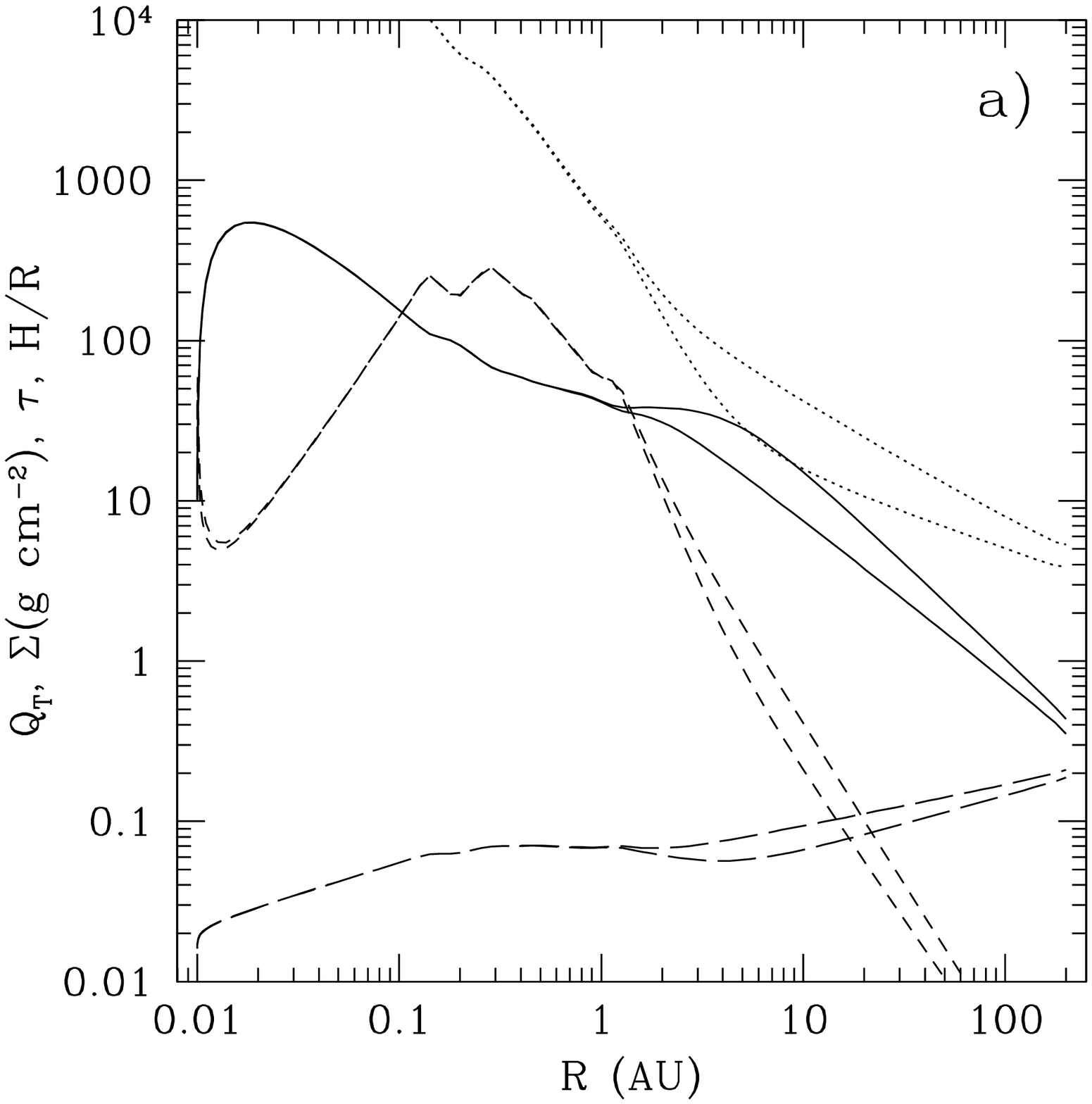}{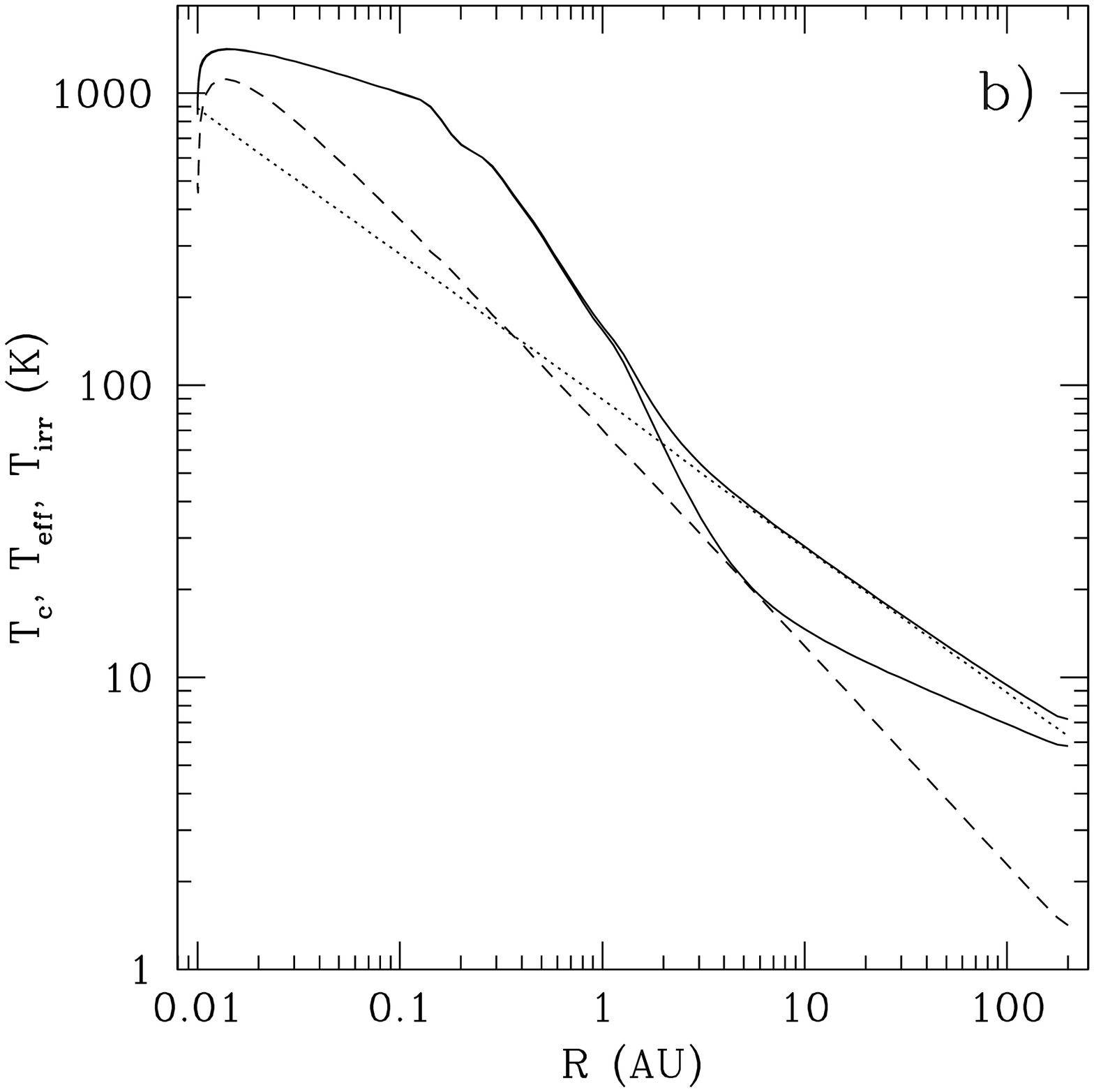}
\caption{Profiles of various quantities in our reference model of a
T-Tauri $\alpha$--disk accreting steadily at a rate $\dot M = 10^{-8}
M_\odot$~yr$^{-1}$, with a viscosity parameter $\alpha =0.02$. (a)
Profiles of surface density, $\Sigma$ (solid line), Rosseland-mean
optical thickness, $\tau$ (short-dashed line), aspect ratio, $h/r$
(long-dashed line) and Toomre $Q_T$ parameter (dotted line) are shown
for both an irradiated and a non-irradiated disk.  At large radii, the
irradiated disk is thicker, less dense and less subject to
self-gravity. (b) Profiles of midplane temperature, $T_c$ (solid
line), and internal effective temperature, $T_{\rm eff}$ (dashed line;
from viscous dissipation only), are shown for the irradiated and
non-irradiated disk models.  The irradiation temperature, $T_{\rm
irr}$, is shown as a dotted line.  At large radii, the irradiated disk
is hotter and it becomes vertically isothermal.
\label{fig:one}}
\end{figure}

Figure~\ref{fig:one}a shows the radial profiles, from $0.01$ to
$200$~AU, of the surface density (solid line), Rosseland-mean optical
thickness (short-dashed line), aspect ratio (long-dashed line) and
Toomre $Q_T$ parameter for self-gravity (dotted line) in the reference
model. For each quantity, a profile with and without disk irradiation
is shown. Note that the profiles very near the inner edge of the
numerical domain are affected by our choice of boundary condition but
they become independent of it beyond $0.02$~AU or so. The disk is
characterized by an aspect ratio rising from $\gsim10^{-2}$ close to
the central star to $\gsim 0.1$ at AU distances and beyond. Disk
self-gravity is at most weak, even for the coldest outer regions, at
$r=200$~AU. The disk makes a transition from an optically-thick regime
to an optically-thin regime at a distance of several AUs from the
central star.

Figure~\ref{fig:one}b shows radial profiles of the midplane
temperature (solid line) and internal effective temperature, arising
from internal viscous dissipation only (dashed line), for the same
reference model with and without irradiation. The transition from an
optically thick to an optically thin regime is again easily identified
at a distance of several AUs from the central star. In the irradiated
model, the outer, optically-thin regions of the disk are vertically
isothermal (the midplane temperature being equal to the irradiation
temperature).

In the outer regions of the non-irradiated model, the midplane
temperature (lower solid line) is significantly in excess of the
effective temperature (dashed line) because of the inefficiency of the
optically-thin gas at radiating the dissipated energy. It is clear
from figures~\ref{fig:one}a and~1b that the effect of disk irradiation
is relatively minor and that it only is important for regions of the
disk beyond 1~AU from the central star. Note also that regions of
opacity transitions in the disk can already be identified in
figure~\ref{fig:one} as regions where the slopes of many of the
profiles shown change. As we will see later, these slope changes have
potentially important effects on the planetary migration process.

Our results for the steady-state structure of a typical T-Tauri
$\alpha$-disk are in good overall agreement with previous comparable
studies (see, e.g., D'Alessio et al. 1998; Papaloizou \& Terquem 1999;
Fromang et al. 2002). In particular, the comparison to D'Alessio et
al. (1998) is quite satisfactory given that these authors treated the
disk structure in considerably more details than we did. These
successful comparisons validate {\it a posteriori} a number of
simplifying assumptions that we made in our treatment of the disk
structure (see \S2).

This success should not, however, eclipse the fact that there remains
considerable theoretical uncertainty regarding the structure of
proto-planetary disks. Perhaps the most important source of
uncertainty comes from our ignorance of the nature of angular momentum
transport (``viscosity'') in the bulk of proto-planetary disks (Gammie
1996; Glassgold, Najita \& Igea 1997; Fromang et al. 2002; Fleming \&
Stone 2003; Matsumura \& Pudritz 2003). Over the past few years, it
has also become clear that the role of a super-heated layer of dust is
important to explain the observational characteristics of T-Tauri
disks (Chiang \& Goldreich 1997), an aspect of the problem that has
simply been ignored in our treatment.

\subsection{Tests of Lindblad Torque Calculations}

Tests of our implementation of the torque density calculations serve
several purposes. We want to recover previously established results,
guarantee numerical convergence and identify the main differences
between the classical 2D treatment of spiral density wave theory and
our simple 3D treatment.

\begin{figure}
\plottwo{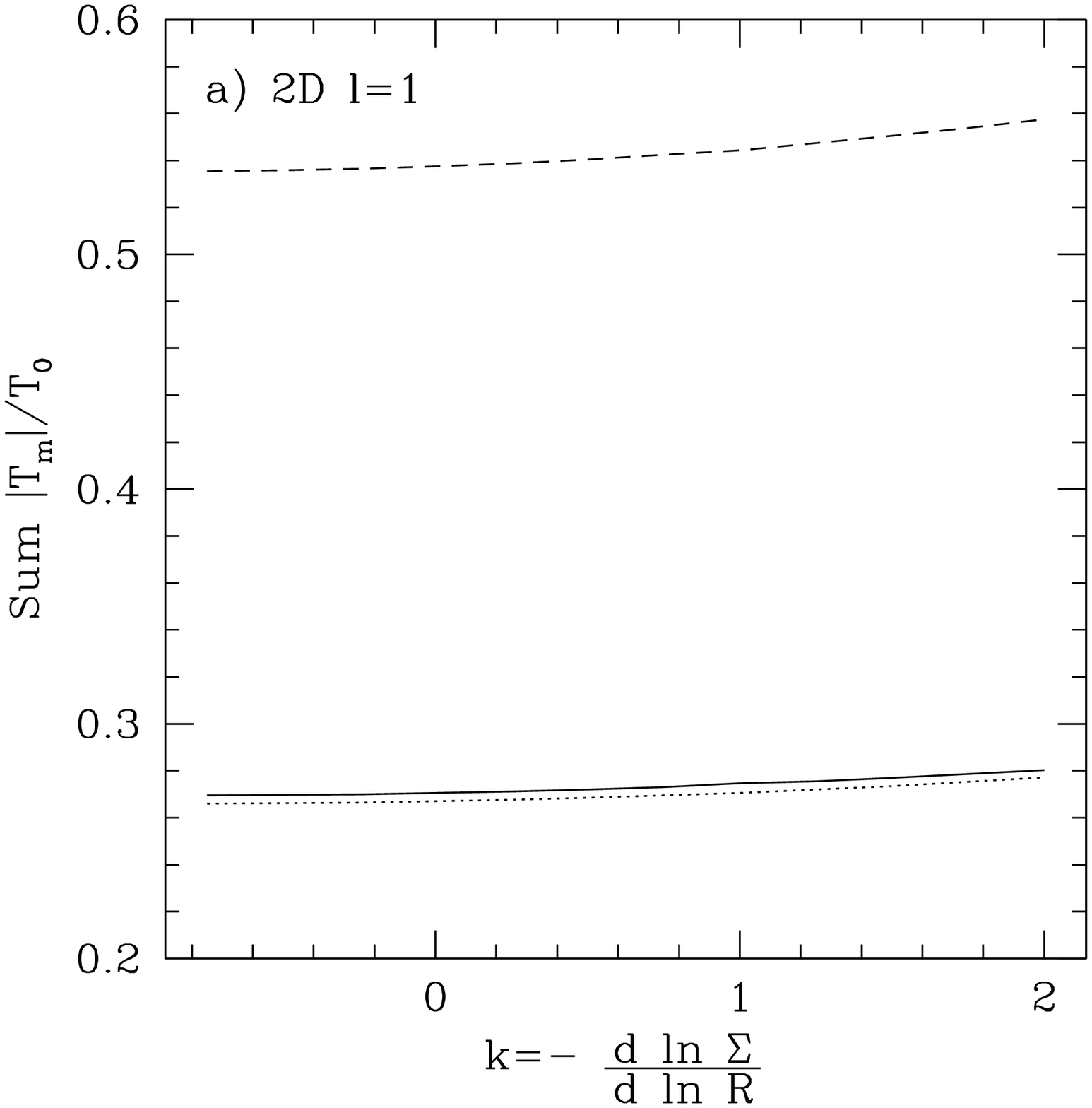}{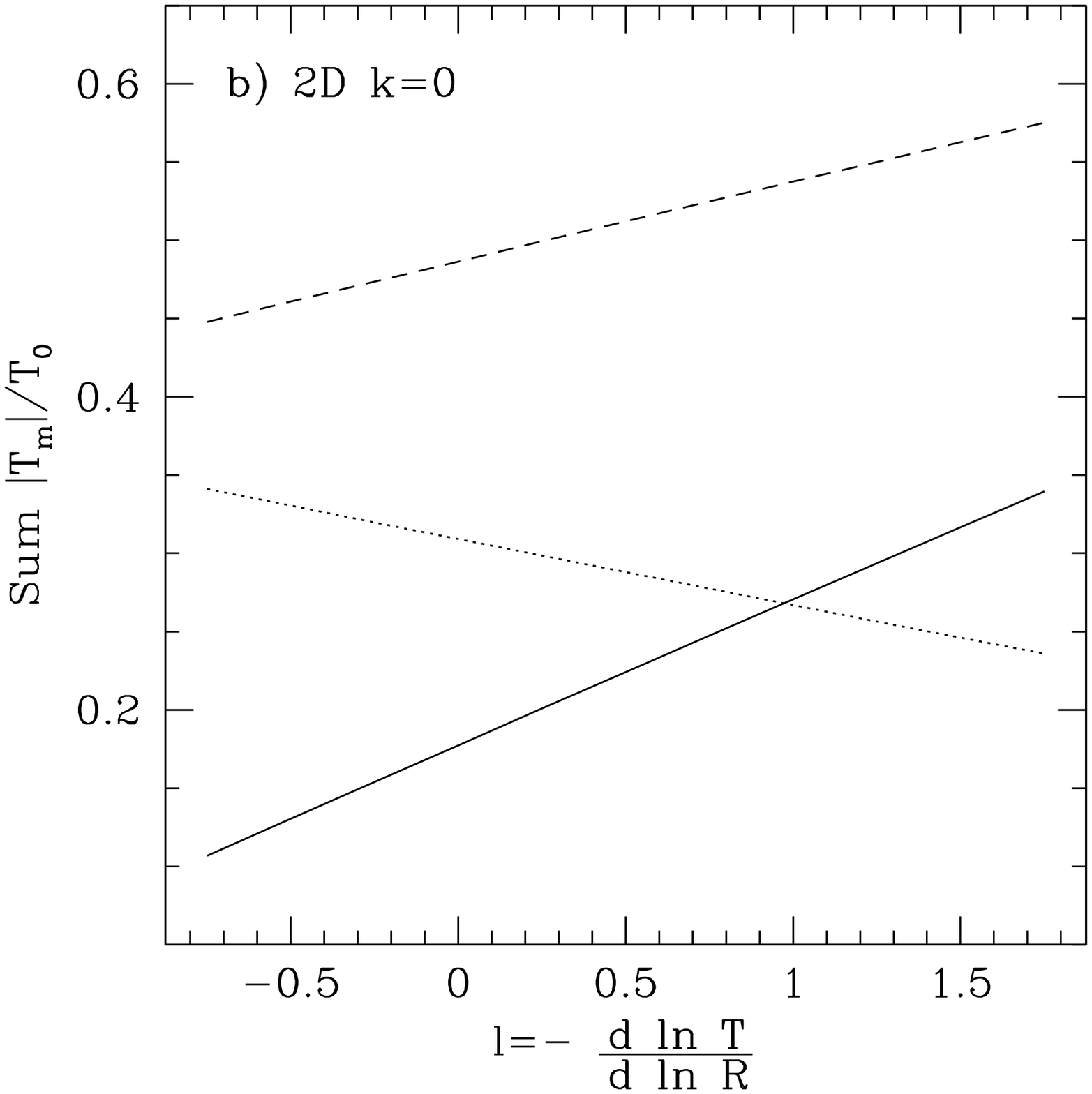}
\caption{Normalized Lindblad torques for a disk with power law radial
profiles of surface density, $\Sigma$, and temperature, $T$, and a
fixed aspect ratio, $h/r=0.07$. The two-dimensional formalism of
spiral density wave theory is adopted here and the results of Ward
(1997a) are recovered with high accuracy (compare his Fig.~3). (a)
Inner, outer and net torques (dotted, dashed and solid lines,
respectively) are shown for a disk with a fixed temperature profile
($l=1$). (b) Same for a disk with a fixed surface density profile
($k=0$).
\label{fig:two}}
\end{figure}

Figure~\ref{fig:two} shows results for our 2D implementation, in a
format such that they can be directly compared to similar results
obtained by Ward (1997a), for a disk with fixed aspect ratio,
$h/r=0.07$. The inner, outer and differential Lindblad torques (or,
equivalently, integrated torque densities) are normalized to the value

\begin{equation}
T_0= \pi \Sigma_p r_p^4 \Omega_p^2 \left( \frac{M_p}{M_*} \right)^2 
\left( \frac{h_p}{r_p} \right)^{-3}, 
\end{equation}
where the subscript $p$ refers to the proto-planet's location.

The background disk is idealized by assuming that the radial profiles
of surface density and midplane temperature are exact power laws, with
indexes

\begin{equation}
k \equiv - \frac{d \ln \Sigma}{d \ln r}, ~~l \equiv - 
\frac{d \ln T_c}{d \ln r}.
\end{equation}

A comparison between our figure~\ref{fig:two} and figure~3 of Ward
(1997a) shows that we recover his results on both the magnitude of
Lindblad torques and their dependence on $k$ and $l$ with high
accuracy. In particular, we confirm that 2D Lindblad torques are
essentially independent of the background profile of surface density
(as measured by $k$).

\begin{figure}
\plottwo{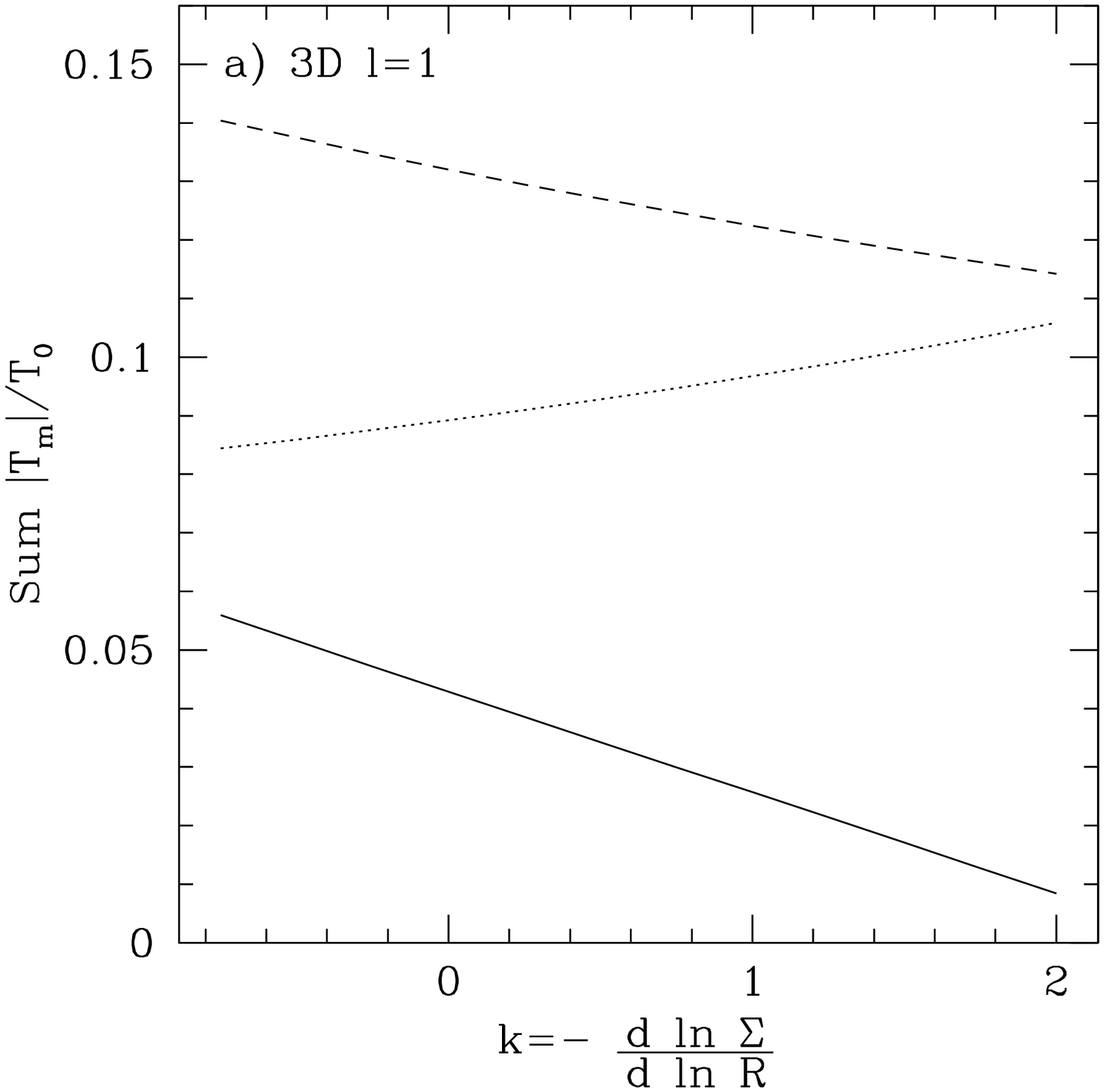}{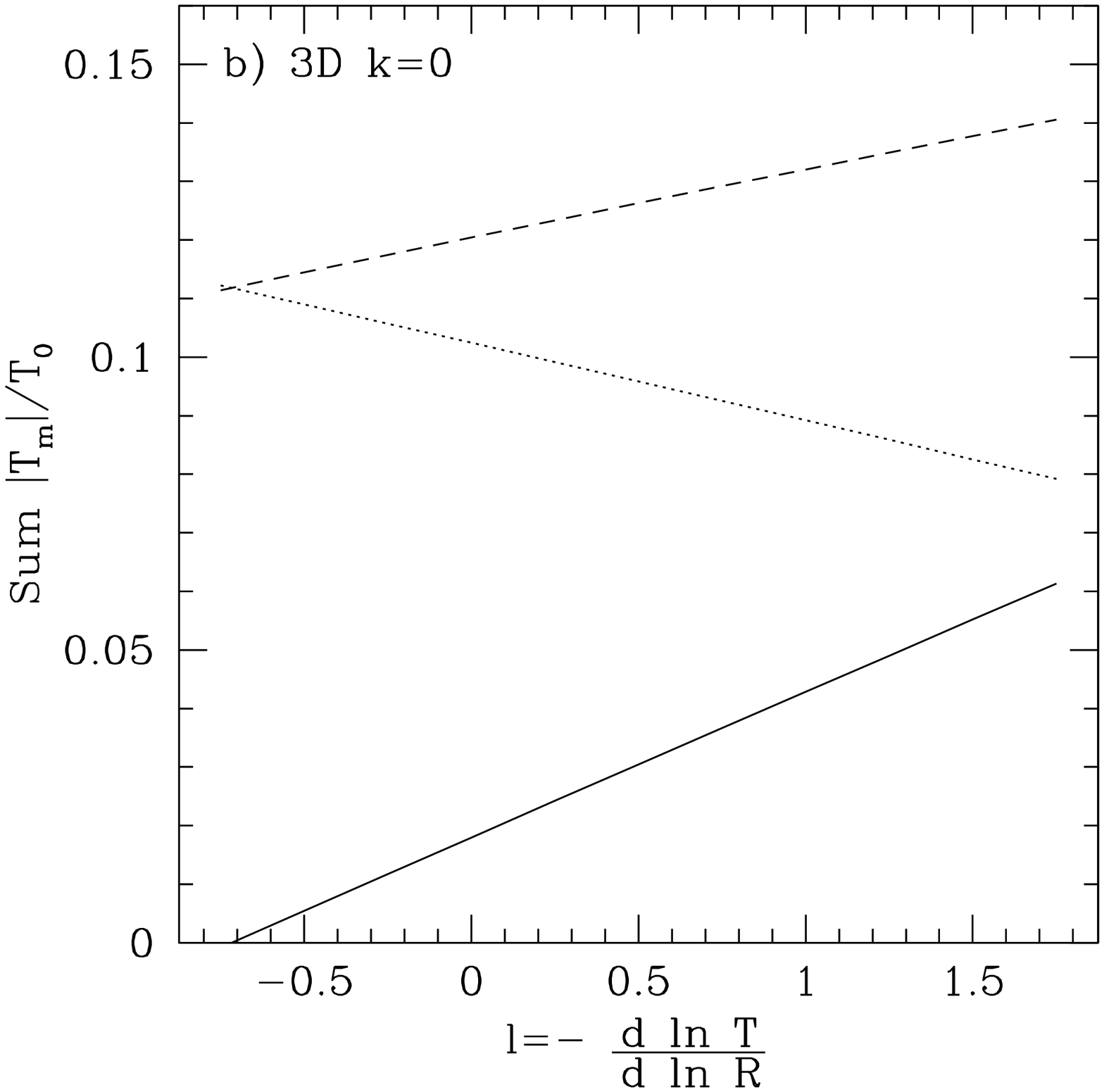}
\caption{Same as Fig.~\ref{fig:two}, but this time adopting a modified
formalism which accounts for three-dimensional effects in the torque
calculation (see text for details). Note the clear reduction in all
torque values and the qualitatively different dependence on the
surface density profile.
\label{fig:three}}
\end{figure}

Figure~\ref{fig:three} shows very different results for our simple 3D
implementation of Lindblad torques. The 3D torques are reduced by a
factor $\sim 3$-$5$ from the 2D values, and their dependence on the
background disk profiles is affected. More precisely, there is now a
clear dependence on the background surface density profile (as
measured by $k$). In fact, the weakness of 3D Lindblad torques and
their dependence on the background disk properties are such that a
sign reversal of the differential Lindblad torque appears possible for
sufficiently steep surface density profiles (large $k$) and
sufficiently shallow midplane temperature profiles (small $l$).

We have confirmed that the results shown in figures~~\ref{fig:two}
and~\ref{fig:three} are unaffected by changes of numerical resolution
in the background disk models, in the range $N=100$ to $1000$. These
tests show adequate numerical convergence of our torque density
calculations based on a sub-grid method (see \S3.2).

\subsection{Migration Rates}

Given a proto-planet of mass $M_p$ embedded in a disk at a radius
$r_p$, the net torque, $\dot J_p$, experienced by the proto-planet is
calculated as a function of the disk properties in the vicinity of
$r_p$. In the present work, we neglect the feedback on the disk
structure caused by the proto-planet's influence. This assumption
should be reasonably accurate in the low-mass limit.  Rather than
following the actual migration with time of the proto-planet in the
gaseous disk, we estimate the local inward
migration time of the proto-planet as

\begin{equation}
T_{\rm mig} \sim \frac{J_p}{- \dot J_p} \simeq \frac{M_p \sqrt{G M_*
r_p}}{- \dot J_p},
\end{equation}

where we have assumed a low-mass proto-planet again in the expression
for the Keplerian angular momentum, $J_p$. By populating the entire
disk with a sufficient number of identical mass proto-planets, we
obtain a radial profile of local migration times. In what follows, we
show how such migration time profiles are affected by variations in
the background disk model. We have found that populating the disk with
200 identical proto-planets, logarithmically spaced from $r_{\rm
in}=0.01$~AU to $r_{\rm out} =200$~AU was enough to obtain accurate
migration time profiles in all cases of interest.

\begin{figure}
\plotone{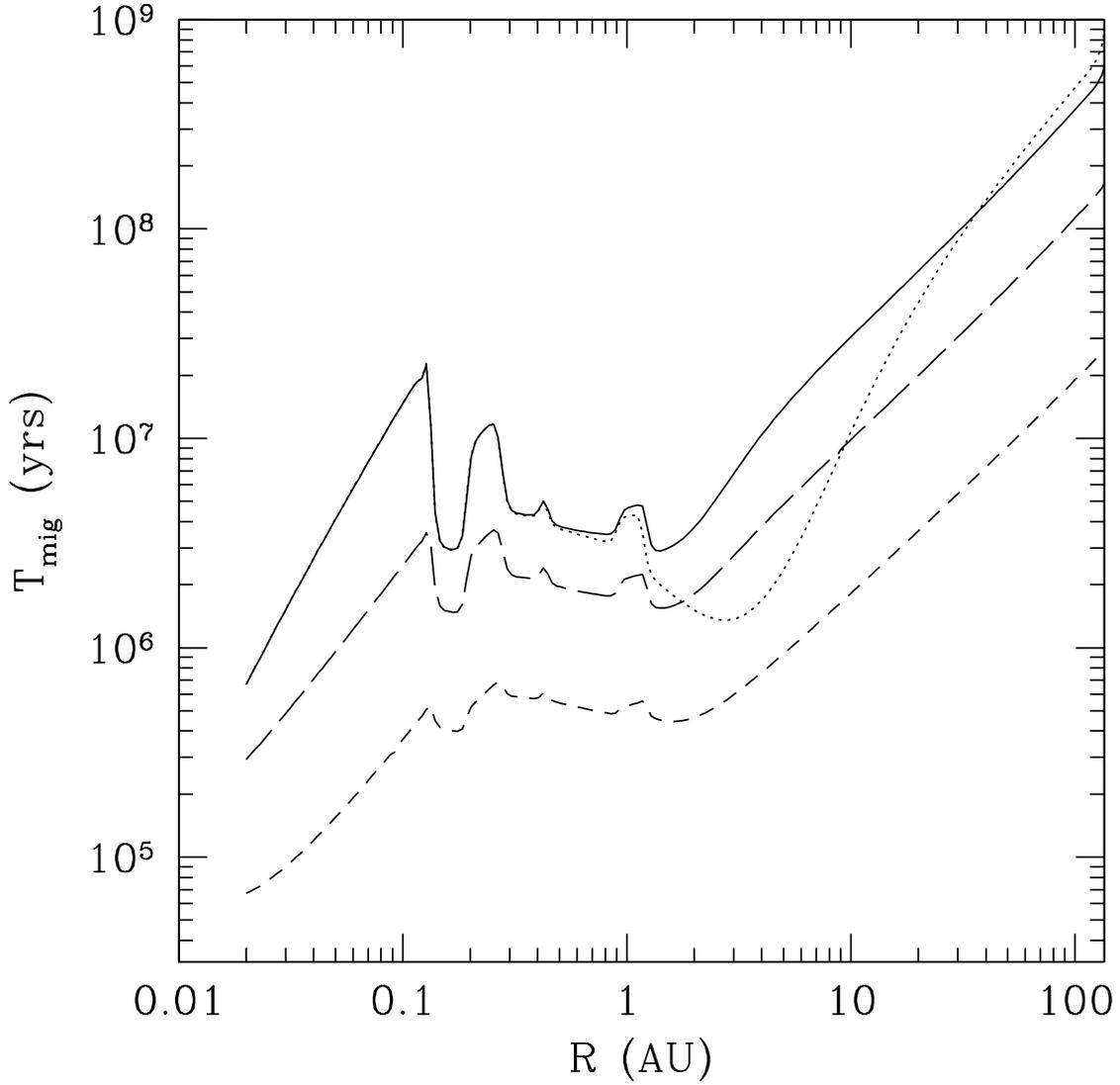}
\caption{Inward migration time of an Earth-mass proto-planet as a
function its location in our reference T-Tauri $\alpha$--disk model,
with $\dot M = 10^{-8} M_\odot$~yr$^{-1}$ and $\alpha = 0.02$. Solid
and short-dashed lines compare the results obtained with the 3D and 2D
formulations, respectively, for an irradiated disk. The dotted line
shows, using the 3D formulation, the results obtained for a
non-irradiated disk model (differences appear only at large radii).
{ The long-dashed line shows the results obtained with a modified
3D formulation, in which the gas is assumed to be vertically extended
over only half the local disk scale height.}  Migration times obtained
with the 3D formulation are significantly longer than those obtained
with the 2D one and they are more sensitive to opacity transitions in
the disk. Regions of maxima of the migration times may be preferred
sites for interactions between proto-planets.
\label{fig:four}}
\end{figure}

Figure~\ref{fig:four} shows radial profiles of inward migration time,
$T_{\rm mig}$, for Earth-mass proto-planets embedded in our reference
T-Tauri disk model, with $\dot M = 10^{-8} M_{\odot}$~yr$^{-1}$ and
$\alpha=0.02$ (see Fig.~\ref{fig:one} for detailed disk
properties). The solid line shows migration times calculated with the
3D formalism, while the short-dashed line shows the same result
according to the standard 2D formalism of spiral density wave
theory. As expected, migration times are significantly longer in the
3D formulation. The dotted line shows, for the 3D formulation, that
the effects of neglecting disk irradiation on the estimated migration
times are significant only at distances $\gsim 1$~AU from the central
star, where the disk structure becomes sensitive to irradiation.  {
In addition, the long-dashed line shows the profile of migration times
obtained with a modified 3D formulation, assuming that the gas is
vertically extended over only half a local disk scale height for the
torque calculation ($h \rightarrow 0.5 h$ in Eq.~[\ref{eq:pot}]). This
confirms that migration times are sensitive to the exact vertical
distribution of gas in the disk.}

The most striking features of figure~\ref{fig:four} are the sudden
variations of $T_{\rm mig}$ with radius, at distances $\lsim 1$~AU
from the central star. Due to a stronger dependence on the background
disk properties (in particular on the surface density profile), radial
variations of $T_{\rm mig}$ are much more pronounced in the 3D
formulation of Lindblad torques than in the 2D one, even if they are
clearly present in both cases. It is also apparent that these
variations of $T_{\rm mig}$ occur at specific locations in the
disk. They are easily identified as the regions where changes of
opacity regime occur in the disk (see the corresponding slope changes
in radial profiles of $\Sigma$, $T_c$ and other quantities in
Fig.~\ref{fig:one}). These radial variations of $T_{\rm mig}$ (or,
equivalently, of the differential Lindblad torque) have their origin
in differences between the inner and outer disk properties when the
proto-planet is located in the close vicinity of an opacity
transition. In these special regions of the disk, with different inner
and outer slopes for $\Sigma$ and $T_c$, differential Lindblad torques
can take values that are quite different from more typical values
adopted in the rest of the disk.

This property of differential Lindblad torques is interesting because
it suggests that narrow regions (e.g., at $0.1$ and $0.25$~AU in
Fig.~\ref{fig:four}) corresponding to local maxima of the migration
times may be the preferred sites of mutual interactions between
proto-planets. Clearly, this possibility and its consequences can only
be investigated by actually following the time-dependent migration of
a population of proto-planets, and accounting for their mutual
gravitational interactions as they approach stalling regions such as
the ones identified in figure~\ref{fig:four}.

Another notable feature of the migration times shown in
figure~\ref{fig:four} is their overall magnitude. In agreement with
previous studies (see, e.g., Fig.~14 of Ward 1997a), we find that the
typical migration time of an Earth-mass proto-planet located at a few
AU distances from the central star is of order a few $10^5$~years,
according to the 2D formalism of spiral density wave theory. As we
already mentioned in \S1, this relatively short timescale has been
recognized as a major difficulty for the standard core accretion
scenario for giant planet formation because it may not allow enough
time for the build up of a substantial gaseous envelope. On the other
hand, the typical migration times obtained with the 3D formulation are
typically one order of magnitude longer than the 2D ones, and regions
in the vicinity of opacity transitions are characterized by migration
times approaching $10^7$~years for an Earth-mass proto-planet. We view
this result as an indication that a proper treatment of 3D effects in
Lindblad torque calculations, together with a detailed background disk
model, may be essential to the resolution (even if only partial) of
this long standing difficulty for the core accretion scenario.

In what follows, we investigate the effect on migration times of
variations in the background disk model. Since the 3D formalism is
arguably the most appropriate (proto-planetary disks do have a finite
thickness) and it is more sensitive than the 2D one to details in the
background disk properties, we chose to show results for migration
times calculated with the 3D formulation only. We also restrict
ourselves to disk models that include the effects of irradiation by
the central star on the disk structure.

\begin{figure}
\plotone{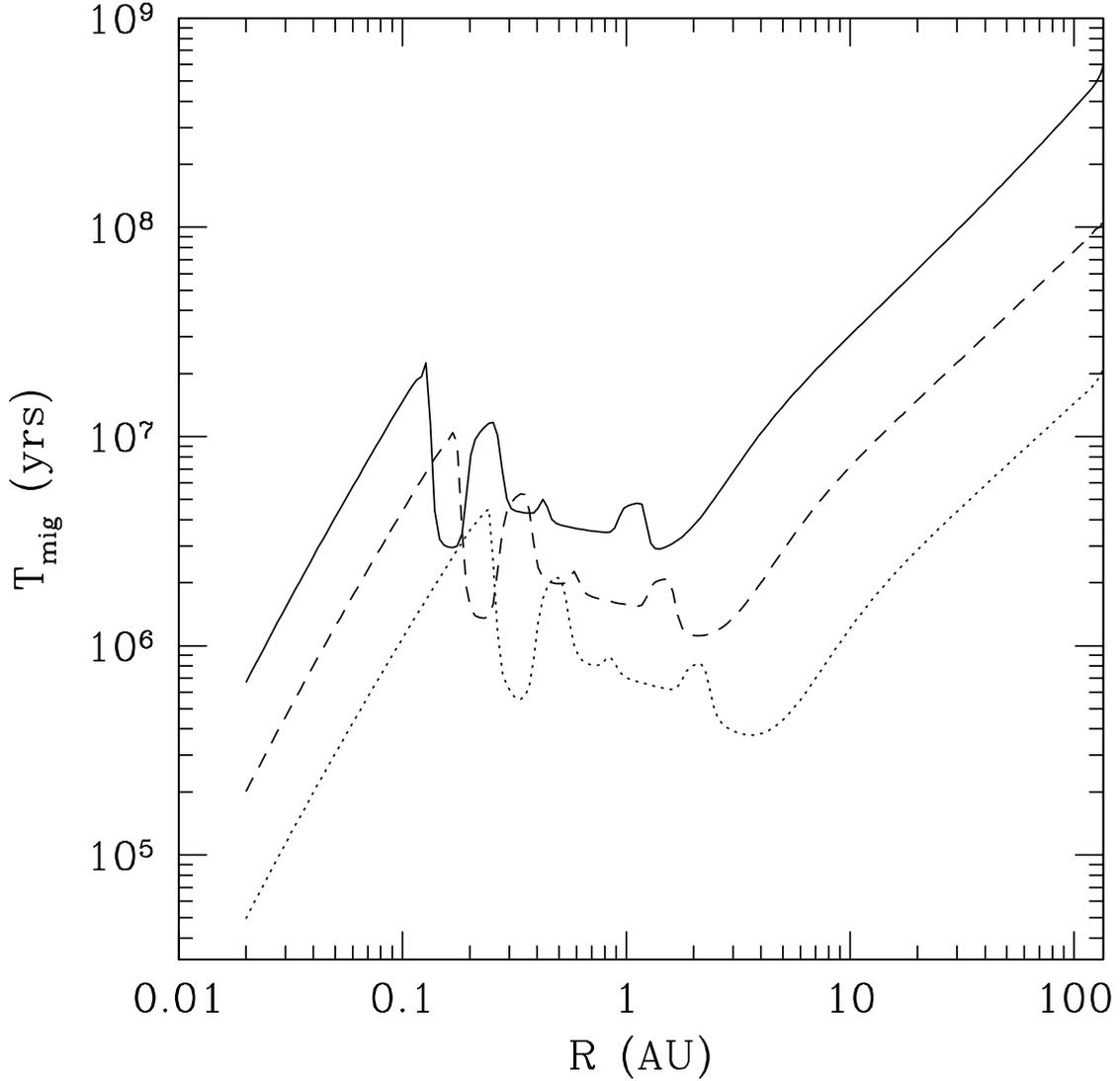}
\caption{Inward migration time of an Earth-mass proto-planet as a
function its location in steady-state T-Tauri $\alpha$--disks with
$\dot M = 10^{-8} M_\odot$~yr$^{-1}$ and different values of the
viscosity parameter, $\alpha$. The 3D formulation of Lindblad torques
is adopted and disk irradiation is included.  Solid, dashed and dotted
lines correspond to $\alpha = 2 \times 10^{-2}$ (reference model), $5
\times 10^{-3}$ and $10^{-3}$, respectively. Decreasing the value of
$\alpha$ increases the disk mass (which reduces migration times) and
increases somewhat the values of radii at which opacity transitions
occur.
\label{fig:five}}
\end{figure}

Figure~\ref{fig:five} shows the effect on migration times of varying
the viscosity parameter $\alpha$ in the disk model. As $\alpha$ is
reduced, the disk mass is increased (at a fixed mass accretion rate
$\dot M =10^{-8} M_{\odot}$~yr$^{-1}$), so is the differential
Lindblad torque, so that migration times are reduced. A secondary
consequence of varying $\alpha$ is to increase the values of the radii
at which opacity transitions occur in the disk, so that the regions of
maximum migration times are shifted outward as well. The relative
amplitude of radial variations of $T_{\rm mig}$ remain comparable even
as $\alpha$ is varied.

\begin{figure}
\plotone{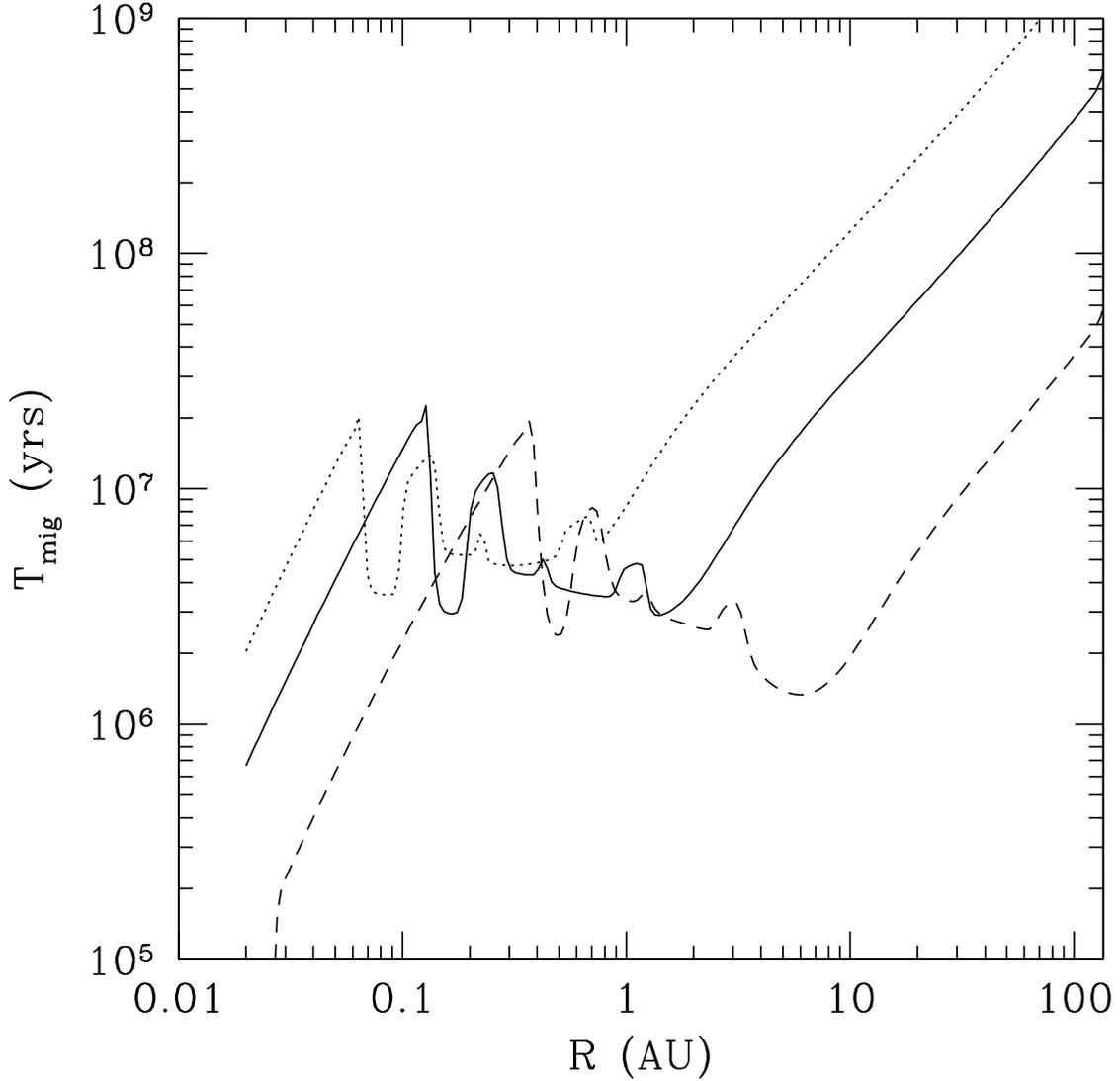}
\caption{Inward migration time of an Earth-mass proto-planet as a
function its location in steady-state T-Tauri $\alpha$--disks with
$\alpha=0.02$ and different values of the mass accretion rate, $\dot
M$. The 3D formulation of Lindblad torques is adopted and disk
irradiation is included.  Solid, dashed and dotted lines correspond to
$\dot M =10^{-8}$ (reference model), $10^{-7}$ and $2.5 \times
10^{-9}$ $M_\odot$~yr$^{-1}$, respectively. Increasing the value of
$\dot M$ increases somewhat the disk mass (which reduces migration
times) and increases the values of radii at which opacity transitions
occur.
\label{fig:six}}
\end{figure}

Figure~\ref{fig:six} illustrates the effect of varying the mass
accretion rate $\dot M$ in a background disk model with the same
viscosity parameter $\alpha=0.02$ as in our reference model. This
time, the overall magnitude of migration times is only slightly
affected (disks with larger $\dot M$ are somewhat more massive), while
the profiles are shifted outward as the mass accretion rate is
increased. {Finally, we note that we have checked directly the linear
dependence of $T_{\rm mig}$ with the proto-planet's mass, as expected
for this migration of ``type I'' (without disk
feedback). Consequently,} even for a proto-planet of mass $M_p=10
M_\oplus$, which in some specific core accretion models is considered
as a typical threshold mass for the onset of runaway envelope
accretion, migration times in some regions (at sub-AU distances) of
our reference T-Tauri disk model are still in excess of $10^6$~years.

\subsection{Conditions for Gap Formation}

One of the assumptions made in our derivation of migration times is
that the feedback torque exerted on the disk by a proto-planet is
negligibly small. While this assumption should be reasonably accurate
in the low-mass limit, we can further quantify it by estimating the
mass above which the feedback torque becomes strong enough for the
proto-planet to open a gap. Following Rafikov (2002; see also Lin \&
Papaloizou 1986 and Ward \& Hourigan 1989), a proto-planet with a
mass in excess of the "viscous mass"

\begin{equation} \label{eq:lpvis}
M_{\rm V}\simeq \left(10.34  \alpha \right)^{1/2} \left(
\frac{h_p}{r_p} \right)^{5/2} M_*
\end{equation} 

generates a torque larger than the disk's own viscous torque and opens
a gap. In addition, an independent condition for gap
formation is that the proto-planet's Hill radius exceed the local disk
scale-height, which translates into a "thermal mass" limit

\begin{equation}
M_{\rm T} \simeq \left( \frac{h_p}{r_p} \right)^3 M_*.
\end{equation}

The local condition for gap formation is presumably given by the
larger of these two masses.

\begin{figure}
\plotone{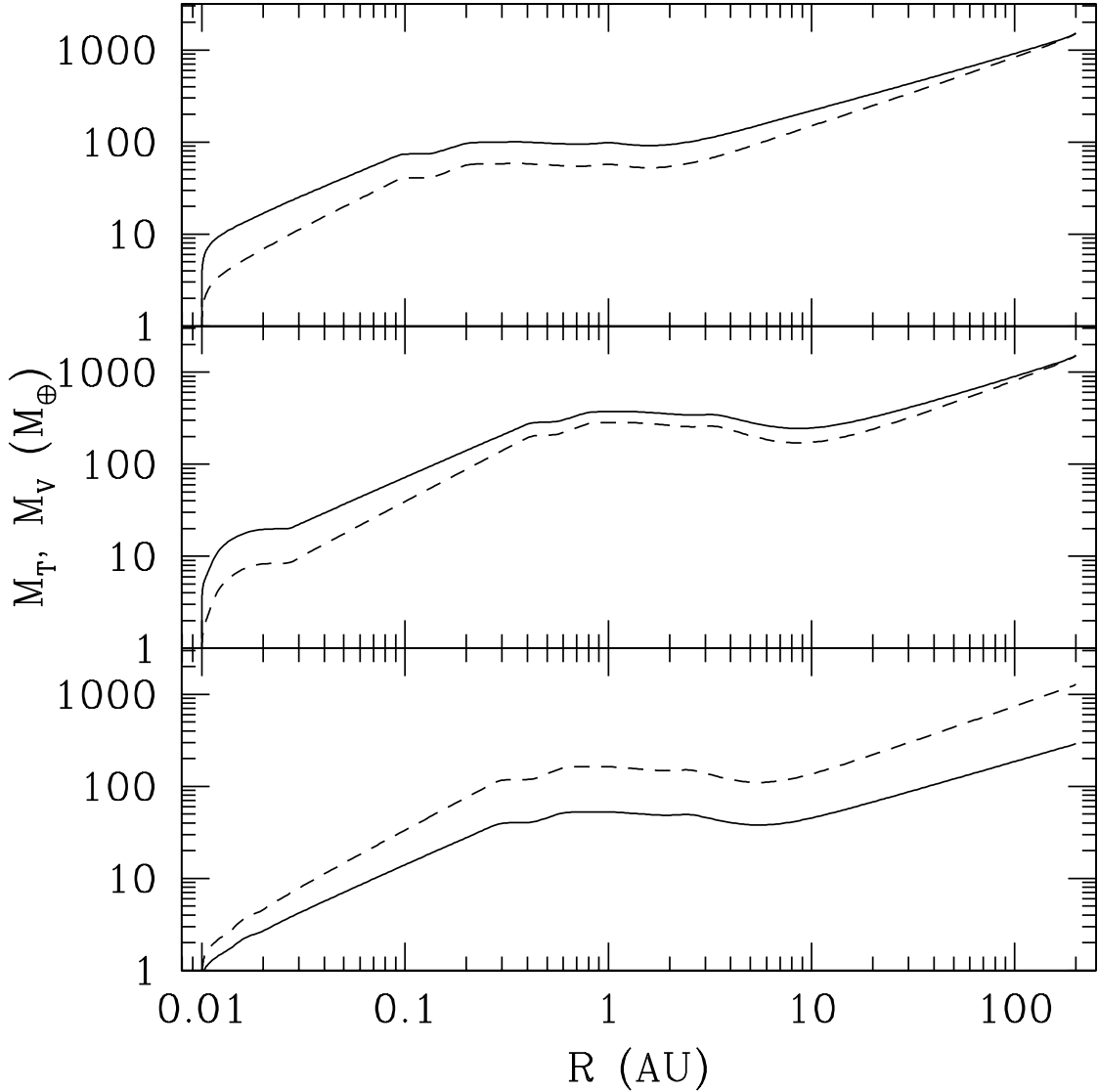}
\caption{Radial profiles of viscous (solid line) and thermal (dotted
line) masses above which a gap is formed (see text for details), in
three of our steady-state T-Tauri $\alpha$-disk models. The upper
panel corresponds to the reference model ($\dot M = 10^{-8}
M_\odot$~yr$^{-1}$ and $\alpha = 0.02$), the middle panel to a
high-accretion rate model ($\dot M = 10^{-7} M_\odot$~yr$^{-1}$ and
$\alpha = 0.02$) and the lower panel to a low $\alpha$ model ($\dot M
= 10^{-8} M_\odot$~yr$^{-1}$ and $\alpha = 10^{-3}$). The viscous mass
is generally the relevant one, except in low $\alpha$ disks. Both
masses are strong functions of radius and they typically exceed $100$
Earth masses at distances $\gsim 0.1$-$0.3$~AU.
\label{fig:eight}}
\end{figure}

Figure~\ref{fig:eight} shows radial profiles of $M_{\rm V}$ and
$M_{\rm T}$ in three representative T-Tauri $\alpha$--disk models for
which we have estimated migration times previously. In each panel, the
solid line corresponds to the viscous mass, $M_{\rm V}$, and the
dashed line to the thermal mass, $M_{\rm T}$. The upper panel shows
the conditions for gap formation in our reference T-Tauri disk model
($\dot M = 10^{-8} M_\odot$~yr$^{-1}$ and $\alpha =0.02$), while the
middle and lower panels show these conditions in a high accretion rate
model ($\dot M = 10^{-7} M_\odot$~yr$^{-1}$) and in a low $\alpha$
model ($\alpha = 10^{-3}$), respectively.
 
In each case, one easily identifies the strong dependence of $M_{\rm
V}$ and $M_{\rm T}$ on the local disk aspect ratio, $h/r$, and how the
location of the plateau at $\sim$AU distances varies somewhat from
model to model. For both the reference and high $\dot M$ models, the
viscous mass, $M_{\rm V}$, determines everywhere the conditions for
gap formation, while the thermal mass, $M_{\rm T}$, is relevant
everywhere in the low $\alpha$ model. Gap formation conditions in
models with intermediate values of $\alpha$ would be determined by
both the viscous and thermal masses, depending on location in the
disk.

Except in the very innermost disk regions, effective gap opening
conditions correspond to masses in excess of $10 M_\oplus$ and reach
values larger than 100 Earth masses typically beyond
$0.1$-$0.3$~AU. In addition, it should be noted that the viscous mass
threshold (Eq.~[\ref{eq:lpvis}]) was determined from a 2D formulation
of spiral density wave theory. Since a more realistic treatment
reveals that 3D Lindblad torques are significantly weaker than 2D
ones, a derivation of $M_{\rm V}$ accounting for the finite thickness
of proto-planetary disks should lead to even larger values of the
viscous mass for gap formation. Therefore, provided one is interested
in proto-planets with masses significantly smaller than the values
shown in figure~\ref{fig:eight}, estimates of local migration times
ignoring disk feedback should be reasonably accurate, { except for
the effects of corotation torques, and except insofar} as the
collective action of many small planets may modify the disk.

\section{Discussion}

Our detailed calculations of Lindblad torques for low-mass
proto-planets embedded in steady-state T-Tauri $\alpha$--disks reveal
strong sensitivities to radial properties of the background disk and
its finite thickness. While we have found these results to be robust
within the set of calculations we performed, several important
limitations should also be emphasized at this point.

{ We have ignored corotation torques for lack of an adequate
parametrization of their effects.  Although Lindblad torques are
usually thought to dominate migration rates (e.g. KP93), the
corotation resonances may be even more sensitive to local conditions
and to feedback than the Lindblad ones, as explained in \S3.3.  Very
likely, corotation torques would alter our quantitative results at
order unity but would reinforce our main qualitative conclusions,
namely, that the migration rate is strongly modulated by opacity
transitions and other radially localized features of $\alpha$ disks.
This will have to be tested in future work.  }

It is unlikely that our steady-state constant--$\alpha$ disk models
are accurate representations of T-Tauri disks in general. Indeed, the
nature of angular momentum transport in proto-planetary disks is not
well understood, even though it determines their structure to a large
extent.  Gammie (1996) observed that MHD turbulence driven by the
magneto-rotational instability (Balbus \& Hawley 1991; 1998) should
operate in the innermost regions ($r \lsim 0.1$~AU) of T-Tauri disks
under standard thermal ionization conditions, but probably not in
their colder, outer regions. He proposed instead that MHD-turbulent
accretion at large radii proceeds only in surface layers that are
sufficiently ionized by cosmic rays. Glassgold et al. (1997) later
refined this layered accretion picture, by taking into account the
ionization due to X-rays from the central star. Since then, it has
become clear that the radial extent of a layered accretion region is
sensitive to details in the disk models (Fromang et al. 2002;
Matsumura \& Pudritz 2003), with weakly-turbulent properties that were
not previously anticipated (Fleming \& Stone 2003).

Absent any contribution to angular momentum transport from
hydrodynamical processes alone (see Balbus 2003 for a discussion),
proto-planetary disks may also have to rely on planetary torques for
their "viscous" evolution (Larson 1990). Goodman \& Rafikov (2001)
studied the propagation and dissipation of spiral density waves in
gaseous disks, within the framework of two-dimensional theory. They
showed that dissipation from wave breaking is quasi-local and they
estimated a magnitude of angular momentum transport $\alpha \sim
10^{-3}$--$10^{-4}$ for a uniform population of Earth-mass
proto-planets embedded in a minimum mass solar nebula. It remains to
be seen what efficiencies of angular momentum transport can be reached
self-consistently in more realistic proto-planetary disk models that
include explicitly the integrated torque feedback from a population of
embedded proto-planets.

Clearly, these large uncertainties on the structure of proto-planetary
disks will have important effects on migration time estimates. We
believe, however, that the identification of opacity transitions in
the disk as regions where migration rates vary rapidly with radius is
not unique to our steady-state $\alpha$-disk models. In the specific
layered accretion model of Gammie (1996), for instance, opacity
transitions cause variations in the slopes of radial solutions, much
as in our models, and this may be a general property of models based
on an $\alpha$-formalism.

What is less clear perhaps is the magnitude of this effect on radial
profiles of migration times. In particular, our models have assumed a
direct dependence of the opacity on the local values of density and
temperature in the disk. In the presence of substantial radial mixing,
for instance due to disk turbulence, these transitions could be made
more gradual and the amplitude of the rapid radial variations of
migration times we have found may be reduced accordingly. The issue of
the turbulent nature of the disk is particularly important in view of
recent results from fully turbulent MHD simulations for embedded
proto-planets (Nelson \& Papaloizou 2003b; see also Laughlin,
Steinacker \& Adams 2003). These simulations show how turbulent
density perturbations can dominate over the specific spiral density
pattern associated with an embedded proto-planet and lead to a largely
random migration process. It remains to be seen whether long-term
migration trends will emerge from such simulations and whether they
will be sensitive to global radial disk properties such as those
exhibited by our $\alpha$--disk models. If not, it may be that our
results on radial variations of migration times will be mostly
relevant to the "dead zones" of layered accretion regions, where the
turbulence is expected to be comparatively much weaker (Fleming \&
Stone 2003).

Not surprisingly, many of the uncertainties we have just highlighted
provide good motivations for additional work on the subject. It would
be interesting to include in our disk models a treatment of layered
accretion similar to that described by Gammie (1997) and study
proto-planet migration in that context. Another important extension of
the present work would be to include the feedback on the disk due to
planetary torques. This would allow us to study more accurately the
conditions for gap formation, as well as the global efficiency of
angular momentum transport in the disk associated with a population of
embedded proto-planets. Ultimately, we may thus be able to study with
some confidence the coupled evolution of a proto-planetary disk and
its population of growing planets.

\section{Conclusion}

We have presented detailed calculations of migration times for
low-mass proto-planets embedded in steady-state T-Tauri
$\alpha$-disks, based on an approximate treatment of two- and three-dimensional
Lindblad torques. We have emphasized the strong sensitivity of local
migration rates to details in the background disk structure and we
have argued that regions of opacity transitions may in general be
characterized by sudden radial variations of the migration
rates. Localized regions with the slowest migration rates may thus be
preferred sites for gravitational interactions between proto-planets.
In some of our disk models, we have found regions with migration times
in excess of $\sim 10^6$~yrs for a $10 M_\oplus$ proto-planet, and
within the framework of the core accretion scenario, this could allow
for significant mass buildup of a gaseous envelope before the
proto-planet is accreted by the central star.

\section*{Acknowledgments}
KM was supported by the Celerity Foundation and the European Community
RTN Program ``The Origin of Planetary Systems'' (PLANETS, contract
number HPRN-CT-2002--00308). JG was supported in part by NASA Origins
grant NAG5-1164.

\clearpage

\clearpage

\clearpage

\clearpage

\clearpage

\clearpage

\clearpage

\end{document}